\documentclass[aps,,showpacs,preprintnumbers,superscriptaddress,prl,twocolumn]{revtex4}
\usepackage{amssymb}
\usepackage{amsfonts}
\usepackage{amsmath}
\usepackage{amsmath}
\usepackage{amssymb}
\usepackage{graphicx}

\begin{document}

\title{Tunneling resonances in quantum dots: Coulomb interaction modifies the width}
\author{Jens K\"{o}nemann}
\affiliation{Institut f\"{u}r Festk\"{o}rperphysik, Universit\"{a}t Hannover,
Appelstrasse 2, D-30167 Hannover, Germany}
\author{Bj\"orn Kubala}
\affiliation{Institut f\"ur Theoretische
Physik III, Ruhr-Universit\"at Bochum, D-44780 Bochum, Germany}
\author{J\"urgen K\"onig}
\affiliation{Institut f\"ur Theoretische
Physik III, Ruhr-Universit\"at Bochum, D-44780 Bochum, Germany}
\author{Rolf J. Haug}
\affiliation{Institut f\"{u}r Festk\"{o}rperphysik, Universit\"{a}t Hannover,
Appelstrasse 2, D-30167 Hannover, Germany}

\date{\today}

\begin{abstract}
Single-electron tunneling through a zero-dimensional state in an asymmetric double-barrier
resonant-tunneling structure is studied. The broadening of steps in the
$I$--$V$~characteristics is found to strongly depend on the polarity of the applied bias
voltage. Based on a qualitative picture for the finite-life-time broadening of the quantum
dot states and a quantitative comparison of the experimental data with a non-equilibrium
transport theory, we identify this polarity dependence as a clear signature of Coulomb
interaction.
\end{abstract}


\maketitle

Single-electron tunneling through zero-dimensional states has been observed
in a wide variety of systems, including metallic islands, lateral quantum dots
in gated semiconductor devices, vertical dots in double-barrier resonant
tunneling structures, and molecular systems such as carbon nanotubes
\cite{review_exp}.
The $I$--$V$~characteristics has the shape of a staircase, in which each step
is associated with the opening of a new transport channel through the system.
Many features observed in the transport measurements
\cite{deshnull,su,equ,thomas,icps} can be explained either within a
single-particle picture for non-interacting electrons or by the orthodox
theory of sequential tunneling \cite{beenakker,averin,glazman} valid for
weak dot-lead tunnel coupling.
This is no longer the case for interaction effects on transport beyond the
weak-tunneling limit.
A famous example is the zero-bias anomaly of Kondo-assisted tunneling
\cite{Kondo_theory,Kondo_exp}.

In this work, we report on a new clear signature of Coulomb interaction beyond weak
tunneling, that is achieved under much less stringent experimental conditions than
required for the Kondo effect to occur. This signature is contained in the width of the
first step of the $I$--$V$~characteristics. We observe that the width strongly depends on
the polarity of the applied bias voltage. This behavior can neither be explained within a
single-particle picture nor by sequential-tunneling theory \cite{beenakker,averin,ralph}.
Due to Coulomb interaction, the finite-life-time broadening of the dot levels becomes
energy dependent. As a consequence, the values for the broadening at the two considered
steps of opposite polarity can differ by up to a factor of two for strongly asymmetric
coupling strengths of the two tunnel barriers. We use the results of a diagrammatic
real-time transport theory \cite{koenig} that includes the above described physics to find
reasonable quantitative agreement with the experimental data.

The experiment was performed with a highly asymmetric double-barrier resonant-tunneling
device grown by molecular beam epitaxy on an n$^+$--type GaAs substrate. An undoped 10~nm
wide GaAs quantum well is sandwiched between 5 and 8~nm thick Al$_{0.3}$Ga$_{0.7}$As
tunneling barriers separated from highly-doped GaAs contacts (Si-doped with $n_\text{Si} =
4\times 10^{17} \text{cm}^{-3}$) by 7~nm thick undoped GaAs spacer layers. The sample was
fabricated as a pillar of $2$~$\mu $m diameter. Two-terminal dc-measurements of the
$I$--$V$ characteristics were performed in a dilution refrigerator at temperatures between
20~mK and 1~K. The studied GaAs quantum well embedded between two AlGaAs barriers can be
viewed as a two-dimensional system with the edges and residual impurities confining the
lateral electron motion and thus forming dots. Tunneling through the energetically lowest
state of the dot, at the energy $E_{0}=33$ meV \cite{prl} and with  a lateral extent of
$10$ nm, produces the lowest resonance peak in the differential conductance $G=dI/dV$.

\begin{figure}
 \center{\includegraphics[width=7.cm]{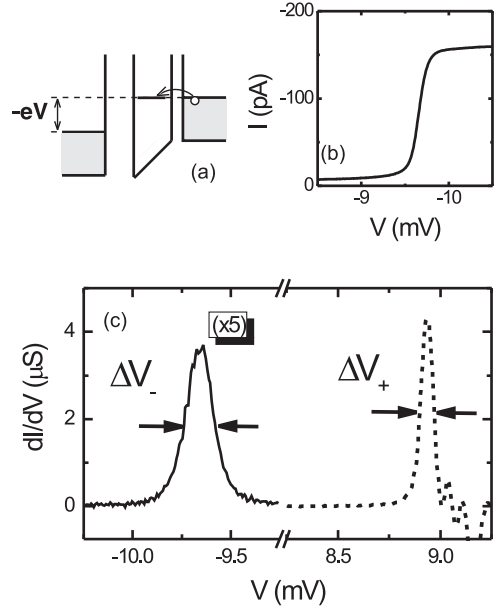}}
\caption{
  (a) Schematic energy diagram of the asymmetric double-barrier device
  under finite bias.
  (b) First current step in negative bias direction.
  (c) Comparison of the full width half maximum (FWHM)--value of the
  differential-conductance peaks for both bias polarities at base temperature
  (T=20~mK).} \label{fig:fig1}
\end{figure}
The system considered is sketched in Fig.~\ref{fig:fig1}(a). Electrons tunnel from the
heavily-doped emitter through the spin-degenerate quantum-dot level embedded inside the
quantum well. The level comes to resonance with the emitter's electro-chemical potential
at a finite bias voltage, indicated by a step in the current, as shown in
Fig.~\ref{fig:fig1}(b).

Coulomb interaction prevents double occupation of the level for the considered range
of bias voltage. This gives rise to charging effects that are differently important for
the two polarities. The bottleneck of transport is provided by the thicker tunnel barrier.
For $V<0$, the ``charging direction'', as sketched in Fig.~\ref{fig:fig1} (a), the dot is
predominantly singly occupied, and, therefore, the two spin channels effectively block
each other. This reduces the current-step height to one half of the value for the opposite
polarity, $V>0$, the ``non-charging direction'', for which the dot is predominantly empty
\cite{averin,glazman,ralph,equ,thomas,icps}. This is well understood within the
sequential-tunneling picture.

The focus of this paper, however, is the finite broadening of the step edge. This
broadening can be measured as the full width at half maximum (FWHM)-value of the
differential conductance ($dI/dV$) peaks associated with the current steps.
Figure~\ref{fig:fig1}(c) shows our experimental low-temperature differential conductance
peaks for both bias polarities. As a result, the width $\Delta V_-=152$ $\mu$V for
negative bias has roughly twice the value than that of $\Delta V_+=71$ $\mu$V. In previous
experiments the resonance width in asymmetric structures has been already studied and  a
polarity-dependence has been observed, but their explanations referred to
polarity-dependent leverage-factors \cite{su} or internal saturation processes \cite{equ}.

 The width of the current
step edge at low temperature reflects the finite-life-time broadening of the
zero-dimensional state due to tunneling in and out of the dot. The golden-rule rates for
an electron tunneling in and out of the dot are given by $\Gamma^+(\omega) =
\sum_{r=\text{L},\text{R}} \Gamma_r f_r(\omega)$ and $\Gamma^-(\omega) =
\sum_{r=\text{L},\text{R}} \Gamma_r [1 - f_r(\omega)]$, respectively. The tunnel-coupling
strength is characterized by the constant $\Gamma_r = 2\pi \nu_r |t_r|^2$, where $\nu_r$
is the density of states in lead $r$, and $t_r$ is the tunneling amplitude, and
$f_r(\omega) = f(\omega-\mu_r)$ is the Fermi function of lead $r$ with electro-chemical
potential $\mu_r$. We note that the presence of strong Coulomb interaction introduces an
asymmetry between tunneling-in and tunneling-out processes: while for an empty dot there
are two possibilities to choose the spin state of the incoming electron, the spin state of
an electron leaving the dot is fixed. It is, therefore, the combination $2\Gamma^+(\omega)
+ \Gamma^-(\omega)$, i.e.,
\begin{equation}
  \sum_{r} \Gamma_r \left[ 1+f_r(\omega) \right] \; ,
\end{equation}
that determines the finite-life-time broadening.
This expression is energy dependent, i.e., it depends on the relative position
of the relevant transport channels to the Fermi energy of the leads.
At low temperature, the broadening probed by the electrons within the
transport window set by the Fermi energies of emitter and collector is either
$2\Gamma_\text{L}+\Gamma_\text{R}$ or $\Gamma_\text{L}+2\Gamma_\text{R}$,
depending on whether the left lead serves as emitter or collector.
This qualitatively explains the polarity-dependent step width.
It is an interaction effect since in the absence of Coulomb interaction the
asymmetry between tunneling-in and tunneling-out processes is lifted by
processes involving double occupancy of the dot, and the finite-life-time
broadening is energy independent, given by $\Gamma_\text{L}+\Gamma_\text{R}$.

\begin{figure}
\centerline{\includegraphics[width=0.7\columnwidth]{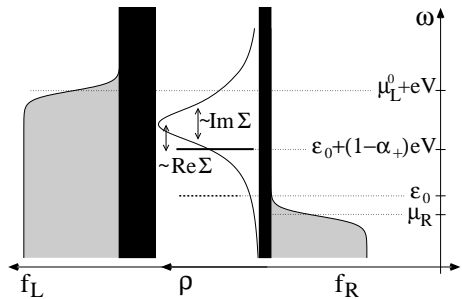}} \caption{
  Sketch of our device in non-charging direction.
  The electronic structure of the dot is given by the spectral density
  $\rho(\omega)$, which is shifted and broadened with respect to the bare
  level $\varepsilon_0+(1-\alpha_+)eV$ (solid line) as reflected in real and
  imaginary part of the self-energy $\Sigma$, respectively.
} \label{fig:fig2}
\end{figure}

For a more quantitative analysis we employ a diagrammatic non-equilibrium
transport theory.
In particular, we make use of the results for the current obtained within
the so-called resonant-tunneling approximation for transport through
a zero-dimensional state in presence of strong Coulomb interaction such
that double occupancy is prohibited.
The technique, the approximation scheme, and the steps of the calculation
are presented in Ref.~\cite{koenig}.
The result for the current is
\begin{equation}
  \label{current2}
  I =\frac{e}{h} \int^{\infty}_{-\infty} d\omega \,
  \frac{2\Gamma_\text{L} \Gamma_\text{R}\,[f_\text{L}(\omega) -
      f_\text{R}(\omega)]}
       { \left[\omega-\varepsilon-\mbox{Re}\,\Sigma(\omega)\right]^2 +
     \left[ \mbox{Im}\,\Sigma(\omega)\right]^2}
  \; ,
\end{equation}
where $\varepsilon$ is the dot level energy [with the appropriate
incorporation of a bias voltage influencing level energy
$\varepsilon=\varepsilon_0 + (1-\alpha_+) eV$ and Fermi-energies of the leads:
$\mu_\text{R}={\mbox const.}<\varepsilon_0,\; \mu_\text{L}=\mu^0_\text{L}+eV$
for the non-charging direction] and the self energy (see Fig.~\ref{fig:fig2})
\begin{equation}
  \label{sigma}
  \Sigma(\omega) = \sum_{r} \frac{\Gamma_r}{2\pi} \int d\omega'
  \frac{1 + f_r(\omega')}{\omega-\omega'+i0^+} \;.
\end{equation}
Evaluating the integral leads to
$\mbox{Re}\,\Sigma(\omega) = \sum_r{\Gamma_r\over 2\pi}
\left[\ln{\left({\beta E_C\over 2\pi}\right)}-\mbox{Re}\,\psi\left({1\over 2}
+i\frac{\beta(\omega-\mu_r)}{2\pi}\right)\right]$ and
\begin{equation}
  \label{Eq. Im sigma dot}
    \mbox{Im}\,\Sigma(\omega) = -\sum_r {\Gamma_r\over 2}[1+f_r(\omega)]\, .
\end{equation}
The real part is weakly dependent on a high-energy cutoff $E_C$ given by the
smaller of the charging energy for double occupancy or the band width of the
leads.
In the imaginary part, we recover the structure of the finite-life-time
broadening as postulated in the qualitative discussion above.
Deep in the Kondo-regime the approximation above is no longer valid. The two spin channels
become independent from each other \cite{Kondo_theory} and  $\mbox{Im}\,\Sigma(\omega) = -\sum_r {\Gamma_r/ 2}\;$.

We remark that for the \emph{step height} the well-known (sequential-tunneling) result
$\Delta I_+= (2e/\hbar) \Gamma_\text{L}\Gamma_\text{R}/(2\Gamma_\text{L}+\Gamma_\text{R})$
for the non-charging direction [and $\text{L}\leftrightarrow \text{R}$  for the charging
direction] is reproduced.

The focus of this paper, however, is on the \emph{width of the step edge}.
At \emph{low temperature}, Eq.~(\ref{current2}) simplifies to
\begin{equation}
  I_+ = \frac{e}{h} \int\limits^{\mu_\text{L}^0+eV}_{\mu_\text{R}} d\omega
  \frac{2\Gamma_\text{L} \Gamma_\text{R}}
  {\left[\omega-\varepsilon - \mbox{Re}\,\Sigma(\omega)\right]^2 +
    \left[ \Gamma_\text{L} + \frac{\Gamma_\text{R}}{2}\right]^2 }
\end{equation}
with $\varepsilon=\varepsilon_0 + (1-\alpha_+) eV$ and $\alpha_+$ being the leverage
factor for positive bias [and $\alpha_-=1-\alpha_+$ being the leverage-factor for negative
bias], denoting the voltage drop over the left barrier. Neglecting the real part of the
self-energy for the moment, we find that the differential conductance as a function of $V$
is a Lorentzian with a FWHM of:
\[
\alpha_+e\Delta V_+ = 2\Gamma_\text{L}+\Gamma_\text{R} \quad \mbox{and} \quad
\alpha_-e\Delta V_- = \Gamma_\text{L}+2\Gamma_\text{R} \; .
\]
For $\alpha_+ \approx \alpha_-$ and strongly asymmetric tunnel-coupling
strengths, $\Gamma_\text{L} \ll \Gamma_\text{R}$, we get the relation
$\Delta V^- \approx 2\Delta V^+$.

At \emph{high temperature} we find for the FWHM $e \Delta V_{\pm} = 3.525 k_{\rm
B}T/\alpha_{\pm} + e\Delta V^0_{\pm}$, i.e. the temperature broadening of the
Fermi-function. As predicted from sequential-tunneling theory~\cite{beenakker} the width
increases linearly with temperature. The constant term  $\Delta V^0_{\pm}$ is of the order
of $\Gamma$, and, in general, also polarity dependent~\cite{remark1}.

\begin{figure}
  \center{\includegraphics[width=0.8\columnwidth]{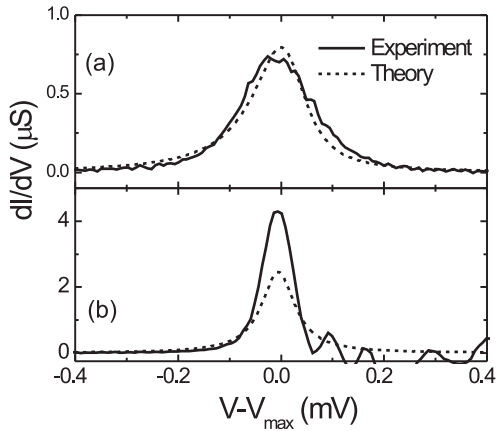}}
  \caption{
    Experimental (solid lines) and theoretical (dashed lines) differential
    conductance for the (a) charging and (b) non-charging
    direction at base temperature (T=20~mK).}
  \label{fig:fig3}
\end{figure}

For a detailed comparison between theory and experiment we need first to determine the
system parameters. The factor $\alpha$ determining the bare level shift with bias voltage
is gained from the linear high-temperature dependence of $\Delta V_{\pm}(T)$ as
$\alpha_+=0.53$ and $\alpha_-=0.5$ so that $\alpha_+ + \alpha_-\approx 1.$ The coupling
constants $\Gamma_{\text{L}/\text{R}}$ could, in principle, both be determined from the
current steps $\Delta I_{\pm}$. For strong asymmetry, however, as is the case here, the
maximum current is limited entirely by the bottleneck of the smaller coupling
$\Gamma_\text{L}$. As a consequence, the step heights only fix $\Gamma_\text{L} = \Delta
I^+ \hbar/(2e)= 2\Delta I^-\hbar/(2e) = 0.64$~$\rm{\mu}$eV.

Considering the results derived above for zero temperature FWHM, we note that this width
is essentially the sum of different couplings and, hence, is dominated by the larger
coupling $\Gamma_\text{R}$. Therefore we can find $\Gamma_\text{R} = (1-\alpha^+)e\Delta
V^+ \approx (1-\alpha^-) \Delta V^-/2\;.$ In fact, we gain better accuracy by fitting the
full $dI/dV$--peak for charging polarity and lowest temperature to pinpoint
$\Gamma_\text{R}= 40$~$\rm{\mu}$eV. The high-energy cut-off $E_C=30$~meV is given by the
bandwidth, i.e. the value of the Fermi energy, being of the same magnitude as the charging
energy.

Figure~\ref{fig:fig3} shows the comparison between experimental data and theoretical
calculations for the differential conductance. For the charging direction (negative-bias),
Fig.~\ref{fig:fig3}(a), we find a good agreement between experiment and theory for both
the resonance width and amplitude. In the non-charging direction (positive bias),
Fig.~\ref{fig:fig3}(b), the peak width is reduced by about a factor of two for both
experiment and theory. The experimental data show some extra features. First there is an
oscillatory fine structure on the positive-voltage side of the resonance attributed to the
fluctuations of the local density of states of the emitter, see, e.g.,
Ref.~\cite{ldos}. Moreover, we observe an enhanced resonance amplitude as compared to the
theoretical calculation. This effect may be related to an additional many-body phenomenon
at the Fermi-edge \cite{int}.

\begin{figure}
 \center{\includegraphics[width=\columnwidth]{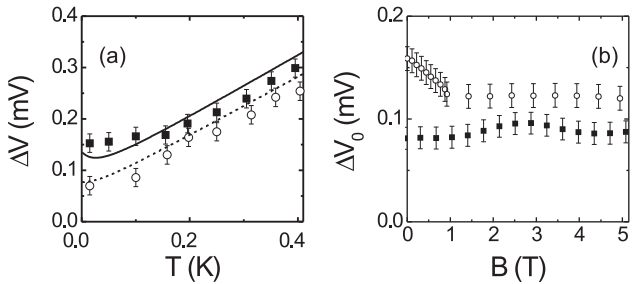}}
 \caption{
   (a) Temperature dependence of the resonance width $\Delta V$ for both
   polarities, open circles $V>0$, filled squares $V<0$, continuous lines
   theoretical simulations.
   (b) Saturation width as function of $B$-field for both polarities.} \label{fig:fig4}
\end{figure}

The temperature dependence of the resonance width in the broad range between $20$~mK and
$400$~mK is shown in Fig.~\ref{fig:fig4}(a). The polarity dependence of the width is
clearly visible. At low temperature experimental and theoretical data nicely match. For
temperatures above 200 mK the width increases linearly with $e\alpha_{\pm}\Delta V_{\pm} =
3.525 k_{\rm B}T + {\cal O}(\Gamma)$. This experimental result is again adequately
reproduced by theory.

Further support for our explanation of the polarity dependence as an interaction effect is
given by the magnetic-field dependence. The data, shown in Fig.~\ref{fig:fig4}(b), reveal
that polarity dependence is strongly reduced by a Zeeman-splitting (on a scale of $\approx
1$~T, which is of the order of the magnitude of the coupling). In this case, only one spin
state contributes to transport, and the system is equivalent to a noninteracting one, for
which theory predicts $\text{Im}\,\Sigma(\omega) = -\sum_r \Gamma_r/2$ and, thus, a
low-temperature width $\alpha_\pm e \Delta V_\pm = \Gamma_\text{L} + \Gamma_\text{R}$
independent of the polarity. This is in accordance with the trend seen in the experimental
data.

The discrepancies remaining between experimental data and theoretical simulation are
consistent with the expected range of accuracy. The main source of experimental errors are
 systematic fluctuations in the current, in particular the local DOS fluctuations
prominent in non-charging direction which become even more pronounced at high
temperatures. This artefact particularly limits the precision to which we can determine
the current step height $\Delta I_+$.

We study the width of the current step of single-electron tunneling through a
zero-dimensional state in a double-barrier resonant tunneling structure and find that it
is strongly polarity dependent. This is interpreted as a clear signature of Coulomb
interaction. The latter introduces an asymmetry between possible tunneling-in and
tunneling-out processes, which gives rise to an energy-dependent finite-life-time
broadening.
The bias-polarity dependent step width, observed in our experiment, where strengths
of the tunnel coupling between dot and the two leads were highly asymmetric,
 can be simulated with reasonable agreement
by applying a diagrammatic non-equilibrium transport theory for interacting quantum dots.

We acknowledge sample growth by A.~F\"orster and H.~L\"uth and discussions with
H.~Schoeller. Financial assistance was granted by BMBF and by DFG via SFB491 and GRK726.


\begin{thebibliography}{99}
\bibitem{remark1}
Neglecting the real part of the self-energy we find for the non-interacting case:
$\alpha\Delta V^0 /(\Gamma^L+\Gamma^R)= 2\sqrt 2/\pi^3 \left[7 \zeta(3)-
2\,\mbox{Re}\,\zeta(3, 1/2+i \ln(1+\sqrt 2)/\pi )\right]=0.6678$ with small deviations
for different polarities in the interacting case.
\end{thebibliography}
\end{document}